\documentclass[paper,11pt]{JHEP}
\usepackage[centertags]{amsmath}
\usepackage{cite}
\usepackage{amsfonts} \usepackage{amssymb} \usepackage{amsthm}
\allowdisplaybreaks[1]
\usepackage{graphicx}
\usepackage{slashed}
\usepackage{latexsym}
\usepackage{bm}
\newcommand{\tr}{\mathrm{tr}\,}
\newcommand{\vol}{\mathrm{vol}}
\newcommand{\dd}{\mathrm{d}}
\newcommand{\eff}{\mathrm{eff}}
\newcommand{\ddef}{\mathrm{def}}
\newcommand{\vev}{\mathrm{vev}}
\newcommand{\trK}{\mathrm{tr_{K}}\,}
\newcommand{\ImP}{\mathrm{Im}\,}
\newcommand{\ReP}{\mathrm{Re}\,}

\title{\boldmath Wilson loops and its correlators with chiral operators in  $\mathcal{N}=2, 4$ SCFT at large $N$}
\author{E.Sysoeva$^{1}$
	\\
	\vskip 0.2cm
	$^1$ Diparimento di Fisica\\
	Universit`a di Roma Tor Vergata\\
	I.N.F.N - sezione di Roma Tor Vergata \\
	Via della Ricerca Scientifica, I-00133 Roma, Italy
\email{sysoeva@roma2.infn.it}
}

\abstract{In this paper we compute the vacuum expectation value of the Wilson loop and its correlators with chiral primary operators in $\mathcal{N}=2, \, 4$ superconformal $U(N)$ gauge theories at large $N$. After localization these quantities can be computed in terms of a deformed $U(N)$ matrix model. The Wilson loops we deal with are in the fundamental and symmetric representations.}


\preprint{ROM2F/2017/04}

\begin{document}

\section{Introduction}

The computation of the vacuum expectation value (vev) of a circular Wilson loop (WL) and the correlators of such WL with chiral primary operators (CPOs) attract interest in the light of the AdS/CFT correspondence\cite{Maldacena:1998im}. Especially interesting  is the possibility to perform exact computations for the vev of a WL in $\mathcal{N}=4$ supersymmetric gauge theories (SYM) exploiting the fact that only planar graphs contribute and that the propagators on the circle are constant \cite{Erickson:2000af}. The problem is thus reduced to the computation of the number of planar graphs, a task which is efficiently performed with the help of the matrix models \cite{Brezin:1977sv, diFrancesco}. From this point of view, the matrix model is just a convenient tool for the computation. A direct computation of the vev of the WL leading to a matrix model was fulfilled in \cite{Pestun:2007rz}. In this paper the partition function of the $\mathcal{N}=2^*$ SYM on $S^4$ was computed and, as a byproduct, it was shown that the instanton corrections vanish when the mass of the hypermultiplet is sent to zero and the $\mathcal{N}=4$ SYM recovers. This establishes a solid link between the computations performed in $\mathcal{N}=4$ SYM and the matrix model. Furthering this line of research, in \cite{Berenstein} the coefficients of the OPE of the WL were computed in the framework of the AdS/CFT correspondence using supergravity and string theory. This computation was later refined with a conjecture for the OPE of correlators of a WL with CPOs of arbitrary conformal dimension \cite{Semenoff}. Finally in \cite{Giombi} D3 and D5 branes were included in the computation. These results were all obtained in the limit of a large number of colors.
\par Following a somewhat different path, the results of \cite{Pestun:2007rz} were extended in \cite{Fucito} by inserting arbitrary powers of the vector superfield into the classical prepotential. The general correlator can then be extracted from a deformed partition function given in terms of a $U(N)$ deformed matrix model replacing the Gaussian matrix model underlying
the $\mathcal{N}=4$ theory. The same type of interacting matrix model arises in the studies of
$\mathcal{N}=2$  theories \cite{Pestun:2007rz, billo}, where now the interactions account for higher loop corrections and mass deformations. In this paper, we study the WL in the deformed matrix model and discuss applications to $\mathcal{N}=2, 4$ superconformal gauge theories.
\par The paper is structured as follows: in Section 2 we study the WL in the fundamental representation, deformed in different ways. In Section 3 the WL in $\mathcal{N}=2$ theory is considered. In Section 4 we calculate the correlators between the WL and CPOs in the fundamental and in symmetric representations.

\section{Deformed matrix models}

We generally follow the notation of \cite{Fucito} and consider deformed $\bf \mathcal{N}=4$ SYM theory on $S^4$ in terms of $\bf \mathcal{N}=2$ SYM theory with one vector multiplet and one adjoint hypermultiplet of matter. The action of the vector multiplet is
\begin{equation}
S_{vec}=\frac{1}{4 \pi^2} \int dx^4 \, d \theta^4 \mathcal{F}_{class}(\Phi) + h.c. \, ,
\end{equation}
where the classical prepotential $\mathcal{F}_{class}$ has the form
\begin{equation} \label{prep}
\mathcal{F}_{class} = \sum_{n=1}^{p}\frac{3 \pi i \tau_n}{n!} \tr \Phi^n \,
\end{equation}
and $\Phi$ is a vector superfield.
\par The WL we are interested in is of the following form
\begin{equation}
W_{\bf R}=\frac{1}{N}\left<\tr_{\bf R} \, e^{\mathcal{C}}\right>_{\vev}=\frac{1}{N} \left<\tr_{\bf R} \, e^{i \, \int_{\mathcal{C}} \left(A_m \dot{x}^m +|\dot{x}|\varphi_1 ds \right)}\right>_{\vev},
\end{equation}
where ${\bf R}$ denotes a representation of the WL, $\varphi_1=\frac{1}{2}(\varphi-\varphi^\dagger)$, $\varphi$ is the scalar of $\Phi$ and $A_m$ are taken to be anti-hermitian matrices. We also choose the contour such that $|\dot{x}|=1$.

As it was shown in \cite{Fucito}, the vev of a WL in the fundamental representation in this case can be written as a $U(N)$ matrix model average
\begin{equation} \label{Wfund}
W=\frac{1}{N}\left<\tr e^{\mathcal{C}}\right>_{\vev}=\frac{1}{N}\left<\tr e^{a}\right>_{\ddef} \, .
\end{equation}
The averages in the matrix model are defined as
\begin{equation} \label{average}
\left<f(a)\right>_{\ddef}=\frac{1}{Z}\int d[a] f(a) \, e^{-NV(a)}=\frac{1}{Z}\int da \, \Delta(a) \, f(a) \, e^{-NV(a)} \, ,
\end{equation}
where
\begin{equation}
Z=\int d[a] \, e^{-N V(a)}=\int da \, \Delta(a) \, e^{-NV(a)}
\end{equation}
with $d[a]$ being the Lebesgue measure in the space of Hermitian $N \times N$ matrices, $da=\prod_{i=1}^{N} da_i$ being the Lebesgue measure in the space of eigenvalues absorbing numerical coefficient irrelevant for the calculation of averages and $\Delta(a)^{\frac{1}{2}}$ being the Vandermonde determinant
\begin{equation}
\Delta(a)=\prod_{u<v=1}^{N}(a_u-a_v)^2 \, .
\end{equation}
The coefficients in the potential $V(a)=\sum_n g_n \tr a^n$ are defined by the parameters $\tau_n$ of the classical prepotential (\ref{prep}) by the rule
\begin{equation}
g_n= \begin{cases}
\frac{4 \pi (-1)^{\frac{n+2}{2}}}{\epsilon^2 N n!} \ImP(\tau_n) \, , \,\,\, n \mid 2
\\
\frac{4 \pi (-1)^{\frac{n+1}{2}}}{\epsilon^2 N n!} \ReP(\tau_n) \, , \,\, n \nmid 2
\end{cases}
\end{equation}
where $\epsilon$ is the reversed sphere radius.
\par In the following subsections we consider different types of the potential $V(a)$.

\subsection{Even potentials}

Let us start with a potential
\begin{equation} \label{pot-def-one-single}
V(a)=g_n \, \tr a^n
\end{equation}
with even $n$.
\par In the large $N$ limit the integrals like (\ref{average}) have been evaluated by saddle point methods \cite{Brezin:1977sv, diFrancesco}.
\par The resolvent defined as
\begin{equation}  \label{res1}
\omega(x)=\frac{1}{N}\left< \tr \frac{1}{x-a}   \right>_{\ddef}=\frac{1}{x}\sum_{k=0}^{\infty}\frac{\omega_k}{x^{k}}
\end{equation}
contains all the averages of the type $\omega_k=\frac{1}{N}\left<\tr a^{k}\right>_{\ddef}$.
\par Under the assumption of a single cut it can be represented in the form
\begin{equation} \label {res}
\omega(x)=\frac{1}{2} V'(x)- Q_{n-2}(x)\sqrt{x^2-b^2} \, ,
\end{equation}
where $Q_{n-2}(x)$ is an even polynomial of order $n-2$. The coefficients of the polynomial as well as the coefficient $b$ are determined by the condition $\omega(x) \approx  \frac{1}{x}$ at large $|x|$. This condition generates a system of $\frac{n}{2}$ linear equations for the coefficients of the polynomial $Q_{n-2}$ and one equation for the coefficient $b$.
\par Solving the system one gets
\begin{equation} \label{res2}
\omega(x)=\frac{n \, g_n \, x^{n-1}}{2} \left( 1-  \sum_{k=0}^{\frac{n}{2}-1}\frac{1}{2^{2k}}\frac{(2k)!}{(k!)^2}	\left(\frac{b}{x} \right)^{2k}\sqrt{1-\frac{b^2}{x^2}} \right) \, ,
\end{equation}
where $b$ is a solution of an equation
\begin{equation} \label{b}
\frac{g_n b^n}{2^n} \frac{n!}{(\frac{n}{2})!(\frac{n}{2}-1)!}-1=0 \, .
\end{equation}
Expanding $\sqrt{1-\frac{b^2}{x^2}}$ in a series one finds the resolvent in a form
\begin{equation} \label{ResSum}
\omega(x) =\frac{1}{x}+\frac{2\, g_n \, \, n!}{2^n(\frac{n}{2}-1)!^2}\sum_{p=1}^{\infty}\frac{1}{2^{2p}}\frac{(2p)! }{(n+2p)p!^2} \frac{ b^{n+2p}}{x^{2p+1}} \, .
\end{equation}
The WL can be written in terms of the resolvent\footnote{The WL can be also found as $
W=\int_{-b}^b \rho(\lambda) e^{\lambda} d \lambda
$, where $\rho(\lambda)$ is density of eigenvalues, which is clearly equivalent to (\ref{WLsum}) since $\omega(z)=\int_{-b}^b \frac{\rho(\lambda)}{z-\lambda}d \lambda$.
}
\begin{equation} \label{WLsum}
W=\frac{1}{2 \pi i}\oint dz \, \omega(z) \, e^{z}dz=\frac{1}{2 \pi i}\oint dz \left(\frac{1}{z}\sum_{k=0}^{\infty}\frac{\omega_{2k}}{z^{2k}}\right) \left(\sum_{t=0}^{\infty} \frac{z^t}{t!}\right)=\sum_{m=0}^{\infty}\frac{\omega_{2k}}{(2k)!} \, .
\end{equation}
From (\ref{res1}), (\ref{ResSum}) and (\ref{WLsum}) one finds
\begin{equation}
W=1 +\frac{2\, g_n \, n!}{2^n(\frac{n}{2}-1)!^2} \sum_{p=1}^{\infty}\frac{1}{2^{2p}}\frac{b^{n+2p}}{(n+2p)p!p!} \, .
\end{equation}
Evaluating the sums one gets
\begin{equation} \label{WLleff}
W=2^{1-n}\frac{g_n \, n!\, }{(\frac{n}{2}-1)!^2} \, b^n  \left(\frac{1}{b}\frac{ \dd }{\dd b}\right)^{\frac{n}{2}-1}  b^{\frac{n}{2}-2} I_{\frac{n}{2}}(b) \, ,
\end{equation}
where $I_n(b)$ is the modified Bessel function of the first kind defined as
\begin{equation*}
I_n (b)= \sum_{m=0}^{\infty} \frac{\left( b/2 \right)^{2m+n}}{m!(m+n)!} \, .
\end{equation*}
Since the resolvent $\omega(x)$ is linear in $g_n$ (except for the implicit dependence on $g_n$ in $b$), the results can be easily generalized to the case of an even potential
\begin{equation} \label{sev-even-pot}
V(a)=\sum_n g_n \tr a^n \, .
\end{equation}
One has simply to sum over $n$.
\begin{equation}
W=\sum_n 2^{1-n} \frac{ \,g_n \, n!\, }{(\frac{n}{2}-1)!^2} \, b^n  \left(\frac{1}{b} \frac{\dd }{\dd b}\right)^{\frac{n}{2}-1}  b^{\frac{n}{2}-2} I_{\frac{n}{2}}(b) 
\end{equation}
with $b$ now satisfying the equation
\begin{equation} \label{b-even}
\sum_n \frac{g_n b^n}{2^n} \frac{n!}{(\frac{n}{2})!(\frac{n}{2}-1)!}-1=0 \, .
\end{equation}
In particular, for the case considered in \cite{Fucito}\footnote{Unlike \cite{Fucito} we choose the parameters of the deformed background as $\epsilon_1=\epsilon_2=2 \pi$, not $\epsilon_1=\epsilon_2=4 \pi$.}
\begin{equation} \label{pot-fuc}
V(a)=\frac{2}{\lambda} \tr a^2 + g_n \tr a^n
\end{equation}
one gets 
\begin{equation} \label{res-fuc}
\omega(x) =\frac{1}{x}+\frac{2 \, b^2}{\lambda \, x}\sum_{p=1}^{\infty}\frac{1}{2^{2p}}\frac{(2p)! }{(n+2p)p!^2} \frac{ b^{2p}}{x^{2p}}+\frac{2\, g_n \, \, n!}{2^n(\frac{n}{2}-1)!^2}\sum_{p=1}^{\infty}\frac{1}{2^{2p}}\frac{(2p)! }{(n+2p)p!^2} \frac{ b^{n+2p}}{x^{2p+1}} \, ,
\end{equation}
\begin{equation} \label{WL-fuc}
W=\frac{2 \, b}{ \lambda}I_{1}(b)+2^{1-n}\frac{ \,g_n \, n!\, }{(\frac{n}{2}-1)!^2} \, b^n  \left(\frac{1}{b} \frac{\dd }{\dd b}\right)^{\frac{n}{2}-1}  b^{\frac{n}{2}-2} I_{\frac{n}{2}}(b) 
\end{equation}
with $b$ being a solution of
\begin{equation} \label{b-fuc}
\frac{g_n b^n}{2^n} \frac{n!}{(\frac{n}{2})!(\frac{n}{2}-1)!}+\frac{b^2}{\lambda}-1=0 \, .
\end{equation}
Although the equation (\ref{b-fuc}) for $b$ cannot be solved for general $n$, it can be treated in perturbation theory in terms of small $g_n$.
\par At the zeroth order one finds the well-known expectation value of the non-deformed WL
\begin{equation} \label{WLnonpert}
W^{(0)}=\frac{2 b}{\lambda}I_1(b), \,\, b^2 = \lambda \, .
\end{equation}
It can be noticed that for the non-deformed WL, $W^{(0)} \sim e^{\sqrt{\lambda}}$ at large $\lambda$ while in the deformed case $W \sim e^{b}$ at large $b$, so $b^2$ can be considered as an effective coupling constant $\lambda_{\eff}$. But unlike the non-deformed case $\lambda_{\eff}$ is limited. In fact (\ref{b-fuc}) implies that
$$
\lambda=\frac{\lambda_{\eff}}{1-\lambda_{\eff}^{\frac{n}{2}}\frac{g_n}{2^{n}} \frac{n!}{(\frac{n}{2})!(\frac{n}{2}-1)!}}
$$
and since $\lambda>0$ one gets a constraint for $\lambda_{\eff}$
\begin{equation*}
 \lambda_{\eff}\leq \lambda_{\eff}^* =4\left(\frac{(\frac{n}{2})!(\frac{n}{2}-1)!}{g \, n!}\right)^{\frac{2}{n}} \, .
\end{equation*}
In the limit of $\lambda \gg \lambda_{\eff}^* $ the WL stops depending on $\lambda$ and turns into a constant $W(\lambda_{\eff}^*)$.
\par Basing on the first two orders of a perturbation expansion one can conjecture the general term of the resolvent
\begin{equation} \label{ResSeries}
\omega(x)=\frac{1}{x}+\sum_{m=1}^{\infty}\sum_{k=0}^{\infty}\frac{(-C(n)g_n\left( \frac{\lambda}{4} \right)^{\frac{n}{2}})^k\left(\frac{\lambda}{4}\right)^m(2m)!(\frac{n}{2}k+m-1)!}{x^{2m+1}m!(m-1)!k!((\frac{n}{2}-1)k+m+1)!} \, ,
\end{equation}
where
$$
C(n)=\frac{(n)!}{(\frac{n}{2})!(\frac{n}{2}-1)!} \, .
$$
For the WL under this conjecture one gets
\begin{equation} \label{WLSeries}
W=1+\sum_{m=1}^{\infty}\sum_{k=0}^{\infty}\frac{(-C(n)g_n\left( \frac{\lambda}{4} \right)^{\frac{n}{2}})^k(\frac{\lambda}{4})^m(\frac{n}{2}k+m-	1)!}{m!(m-1)!k!((\frac{n}{2}-1)k+m+1)!} \, .
\end{equation}
For $n=4,\, 6,\, 8$ the equation (\ref{b-fuc}) for $b$ can be solved exactly. Substituting $b$ into (\ref{res-fuc}), (\ref{WL-fuc}) one finds the result coinciding with (\ref{ResSeries}), (\ref{WLSeries}). For larger $n$ (\ref{b-fuc}) cannot be solved for algebraic reasons.

\subsection{Potentials with odd terms}

In the case of the potential containing both odd and even deformations
\begin{equation} \label{any-pot}
V(a)=\sum_{l=1}^{\infty} g_{l} \,\tr a^{l}
\end{equation}
the resolvent has an asymmetric cut
\begin{equation}
\omega(x) = \frac{1}{2}V'(x)-\sum_{l=1}^{\infty}P_{l-2}(x-c)\sqrt{(x-c)^2-b^2} \, ,
\end{equation}
where the polynomial $P_{l-2}$ contains now both even and odd powers of $(x-c)$ and $c$ stands for the center of the cut. Demanding the resolvent to behave as $\frac{1}{x}$ at large $|x|$ one gets two separate systems of linear equations for the coefficients of $P_{l-2}$ with odd and even powers of $(x-c)$, with the same invertible matrices as in the previous case. This requirement also leads to the equations
\begin{equation} \label{b-odd}
\sum_{l=2}^{\infty} g_l c^{l-2} b^2 \sum_{k=0}^{\left[ \frac{l}{2}\right]-1}\frac{1}{2^{2k+2}}\frac{l!}{(l-2-2k)!}\frac{(b/c)^{2k}}{k!(k+1)!}-1=0 \, ,
\end{equation}
\begin{equation}
\sum_{l=1}^{\infty}g_l c^{l-1} \sum_{k=0}^{\left[\frac{l+1}{2}\right] -1} \frac{1}{2^{2k}}\frac{l!}{(l-1-2k)!}\frac{(b/c)^{2k}}{k!^2}=0 \, .
\end{equation}
 If there are only even powers in the deformation then $c=0$ and the second equation is missing (but $c$ can be equal to $0$ even if there are odd deformations in the potential).
\par The resolvent can now be written as
\begin{eqnarray} \label{res-arb-def}
\omega(x) &=& \frac{1}{x-c} + \sum_{l=2}^{\infty}\sum_{k=1}^{\infty}\frac{1}{2^{2k+2}}\frac{ g_l c^{l-2} b \,l!}{(l-2)!} \frac{(2k)! }{(k+1)! k!} \frac{b^{2k+1}}{(x-c)^{2k+1}} +
\nonumber \\
&+&\sum_{l=3}^{\infty}\sum_{k=0}^{\left[\frac{l+1}{2}\right]-2}\frac{1}{2^{2k+2}}\frac{g_l c^{l-3} b^2 l!}{(l-2k-3)!}  \frac{(b/c)^{2k}}{k!(k+1)!}\sum_{p=1}^{\infty}\frac{1}{2^{2p+1}}\frac{(2p)! }{(k+1+p) p!^2}\nonumber \\
&&\left(\frac{x-c}{b}+\frac{1}{2}\frac{b}{c}\frac{(k+1+p)}{(k+2+p)} \right) \frac{b^{2p+1}}{(x-c)^{2p+1}} \, .
\label{res-arb-def}\end{eqnarray}
For the WL one gets
\begin{equation} \label{WL-arb-def}
W=\frac{e^c}{2 \pi i}\oint dz \, \omega(z) e^{z-c}dz=\frac{1}{2} e^{c}\sum_{l=2}^{\infty} \frac{g_l c^{l-2} b\,l!}{(l-2)!} I_1(b)+
\end{equation}
$$
+\frac{1}{4} e^{c} \sum_{l=3}^{\infty}\sum_{k=0}^{\left[ \frac{l+1}{2} \right]-2} \frac{ g_l c^{l-3}b^2  \, l!}{(l-3-2k)!}\frac{(b/c)^{2k}}{k!(k+1)!} \left( 1 +\frac{1}{2}\frac{b}{c}\frac{(l-3-2k)}{(k+1)}\frac{\dd}{\dd b}\right)\left(\frac{1}{b}\frac{\dd}{\dd b} \right)^{k} b^{k} I_{k+2}(b) \, .
$$

\subsection{Double-trace potential terms}

Let us now consider a potential involving double trace terms
\begin{equation} \label{pot-def-one-double}
V(a)=\frac{2}{\lambda} \tr a^2 + \frac{g}{N} \tr a^n \tr a^m
\end{equation}
which will be used in the Section 3.
\par Considering the averages in this theory as integrals over the eigenvalues (\ref{average}) one gets an equation for the saddle point \cite{Brezin:1977sv}
\begin{equation}
2\sum_{i \neq j =1}^{N}\frac{1}{a_i-a_j}-4\frac{N}{\lambda}a_i-g \, n \, a_i^{n-1}\, \sum_{j=1}^N a_j^m+g \, m \, a_i^{m-1}\, \sum_{j=1}^N a_j^n=0
\end{equation}
which coincides with an equation for the saddle point in the effective potential
\begin{equation} \label{Veff}
V_{\eff}=\frac{2}{\lambda} \tr a^2 + g \, \omega_m \tr a^n + g \, \omega_n \tr a^m
\end{equation}
with $\omega_k$ being the coefficients in the series expansion of the resolvent (\ref{res1}).
\par Thereby the potential (\ref{pot-def-one-double}) can be replaced by an effective potential (\ref{Veff}) and the results from the previous section can be applied also to this case.
\par An explicit form of equations for the coefficients $\omega_m, \, \omega_n$ in the simplest case of even $m, \, n$ follows directly from the resolvent
\begin{equation} \label{omm}
2^m\frac{\omega_m}{b^m}=\frac{b^2}{\lambda}h(m)+\frac{g \, \omega_n b^{m}}{2^{m}} f(m,m)+\frac{g \, \omega_m b^{n}}{2^{n}} f(n,m) \, ,
\end{equation}
\begin{equation} \label{omn}
2^n \frac{\omega_n}{b^n}=\frac{b^2}{\lambda}h(n)+\frac{g \, \omega_n b^{m} }{2^{m}}f(m,n)+\frac{g \, \omega_m b^{m} }{2^{n}} f(n,n) \, ,
\end{equation}
where
\begin{equation*}
h(n)=\frac{n!}{(\frac{n}{2}+1)!(\frac{n}{2})!}, \, \,f(n,m)=\frac{n^2}{2(m+n)}\frac{n!}{\frac{n}{2}!\frac{n}{2}!}\frac{m!}{\frac{m}{2}!\frac{m}{2}!} \, .
\end{equation*}
If $m$ or $n$ is odd, one should be careful since $\omega_k$ is a coefficient in the expansion of the resolvent (\ref{res-arb-def}) in terms of $\frac{1}{x}$, not in terms of $\frac{1}{x-c}$.
\par It can be noticed that there is a difference in the diagrammatic interpretation of a WL in the deformed potentials (\ref{any-pot}) and (\ref{pot-def-one-double}). In the first case every cycle brings a factor $N$, every propagator brings a factor $N^{-1}$ and every vertex brings again a factor $N$, so the total power of $N$ is equal to the Euler characteristic of the surfaces, i.e. all planar diagrams make a contribution in the deformed WL. In the second case every cycle still brings $N$ and every propagator brings $N^{-1}$ while the vertices don't bring a factor $N$ anymore. Therefore not all of the planar diagrams make a contribution. Indeed, in every double vertex appearing in the perturbative treatment of the potential (\ref{pot-def-one-double}) only one vertex can be connected with the WL, while the second one should belong to a part separated from the WL and can be interpreted as an effective factor $\omega_k$ in the first vertex. This fact is reflected by a replacement of the potential by the effective one.

\subsection{Examples}

\begin{itemize}

\item{$V(a)=\frac{2}{\lambda}+ g_4 \, \tr a^4$}
\par This potential was considered in \cite{Fucito}.
\par The WL in this case is
\begin{equation}
W=\frac{2 \, b}{ \lambda}I_{1}(b)+3 \, g_4 \, b^3   \frac{\dd }{\dd b} I_{2}(b) \, ,
\end{equation}
where $b$ is a solution of the equation
\begin{equation}
\frac{3}{4} g_4 b^4 +\frac{b^2}{\lambda}-1=0, \,\, b^2=\frac{-1+\sqrt{1+3 g_4 \lambda^2}}{\frac{3}{2} g_4 \lambda} \, .
\end{equation}
The result coincides with \cite{Fucito}.

\item{$V(a)=\frac{2}{\lambda}+ g_3 \, \tr a^3$}
\par In this case the WL is given by
\begin{equation} \label{WL-arb-def}
W=e^{c}\frac{2 \, b}{ \lambda}I_{1}(b)+3 \, e^{c} g_3 \,  b \, \left( c \, I_1(b)+ \frac{1}{2} \, I_{2}(b)\right) \, ,
\end{equation}
where $c$ is
\begin{equation}
c= \frac{2}{3 \ g_3 \, b^2}\left(1-\frac{b^2}{\lambda}\right)
\end{equation}
and $b^2$ is the real positive solution of the cubic equation
\begin{equation}
9 \, b^6 g_3^2-\frac{b^4}{\lambda^2}+1=0 \, .
\end{equation}

\item{$V(a)=\frac{2}{\lambda}+ g \, \tr a^3 \, \tr a^3$}
\par The corresponding effective potential is
\begin{equation}
V_{\eff}=\frac{2}{\lambda}+ 2 \, \omega_3 \,g \, \tr a^3 \, .
\end{equation}
Re-expanding the resolvent (\ref{res-arb-def}) in terms of $\frac{1}{x}$ one finds the equation for $\omega_3$
\begin{equation}
\frac{\omega_3}{b^2}=\frac{b^2 c}{\lambda}\left( \frac{3}{4}+\frac{c^2}{b^2} \right)+\frac{3}{16} b^4 g_3 \, \omega_3\left( 1+18\frac{c^2}{b^2} \right) \, .
\end{equation}
Together with the equation for $c$
\begin{equation}
\frac{c}{\lambda}+\frac{3}{4} \, b^2 \, g_3 \omega_3+\frac{3}{2} c^2 g_3 \omega_3=0
\end{equation}
it gives the simple solution $c=0, \, \omega_3=0$, so the effective potential does not contain a deformed term at all.
\par One could immediately see that $\omega_3=0$ in this case since the initial potential is even, so all the averages of odd functions of $a$ disappear including
\begin{equation}
\omega_{2k+1}=\frac{1}{N}\left< \tr a^{2k+1}\right>_{\ddef}=\frac{1}{N}\int d[a] \, \tr a^{2k+1} e^{-N \, V(a)}=0 \, .
\end{equation}
Therefore the WL is just
\begin{equation}
W=\frac{2b}{\lambda}I_1(b), \,\, b^2=\lambda \, .
\end{equation}

\item{$V(a)=\frac{2}{\lambda}+ g \, \tr a^2 \, \tr a^4$}
\par The effective potential in this case is
\begin{equation}
V_{\eff}=\frac{2}{\lambda} + \omega_4 \,g \, \tr a^2 + \omega_2 \,g \, \tr a^4 \, .
\end{equation}
The coefficients $\omega_2, \, \omega_4$ are given by the equations (\ref{omm}), (\ref{omn})
\begin{equation}
 \frac{\omega_2}{b^2}=\frac{4}{16- b^6 g}, \,\,\frac{\omega_4}{b^4}= \frac{2+\frac{1}{16} \, b^6 g}{16- b^6 g} \, .
\end{equation}
The WL is
\begin{equation}
W=\frac{2 \, b}{ \lambda}I_{1}(b)+ g \, b^5 \, \frac{2+\frac{1}{16} \, b^6 g}{16- \, b^6 g} \,  I_{1}(b)+12\, g \, b^5 \frac{1}{16- b^6 g}  \frac{\dd }{\dd b} I_{2}(b)
\end{equation}
and $b$ is the real positive root of the equation
\begin{equation}
 \frac{1}{512} \, b^{12} \, g^2 \, \lambda- \frac{1}{16} \, b^8 \, g+\frac{5}{16}\, b^6 \, g \, \lambda+b^2-\lambda=0 \, .
\end{equation}

\end{itemize}

\section{Wilson loop in the $\mathcal{N}=2$ theory}

As it is shown in \cite{billo}\footnote{In \cite{billo} the parameters of deformed background were chosen as $\epsilon_1=\epsilon_2=1$ and we stick to our previous choice $\epsilon_1=\epsilon_2=2 \pi$.} the partition function of the $\mathcal{N}=2$  super Yang-Mills theory with U(N) gauge group and $N_f$ massless fundamental matter multiplets defined on a four-sphere $S_4$ in the weak-coupling limit $g_{YM}=\frac{\sqrt{\lambda}}{\sqrt{N}} \ll 1$ can be written as
\begin{equation} \label{ZN2}
Z_{\mathcal{N}=2}=\int d[a] e^{-\frac{N}{2 \lambda} \tr a^2}e^{-S(a)} \, .
\end{equation}
Here $a$ is a Hermitian $N \times N$ matrices and
\begin{equation}
S(a)=-2 \sum_{u<v=1}^{N} \log H \left(\frac{i \, (a_{u}-a_{v})}{2 \pi}\right)+N_f\sum_{u=1}^{N} \log H\left(\frac{i \, a_{u}}{2 \pi}\right) \, ,
\end{equation}
where
\begin{equation}
\log H(x)=-(1+\gamma)x^2-\sum_{i=2}^{\infty}\zeta(2i-1)\frac{x^{2i}}{i}
\end{equation}
with $\zeta(i)$ being the Riemann zeta-function and $\gamma$ being the Euler-Mascheroni constant.
\par The vev of the circular WL in this theory is
\begin{equation}
W_{\mathcal{N}=2}=\frac{1}{Z_{\mathcal{N}=2}} \int d[a] \tr e^{a} \, e^{-\frac{N}{2 \lambda} \tr a^2} \, e^{-S(a)} \, ,
 \end{equation}
where $S(a)$ can be written as a sum $S(a)=\sum_{i=n}^{\infty} S_{2n}$ of homogeneous polynomials $S_{2i}(a)$ of order $2i$.
\begin{equation}
S_2(a)=-\frac{(1+\gamma)}{(2 \pi)^2}\left( (2N-N_f)\tr a^2  - 2 (\tr a)^2\right) \, ,
\end{equation}
\begin{equation}
S_{2n}(a)=(-1)^n\frac{\zeta(2n-1)}{n \, (2 \pi)^{2n}}\left((2N-N_f)\tr a^{2n} + \sum_{k=1}^{2n-1} C_{2n}^k \tr a^k \tr a^{2n-k}\right) \, ,
\end{equation}
where $C_n^k$ are the binomial coefficients.
\par Let us consider a conformal case with $N_f=2N$.
\par $S(a)$ can be interpreted as a deformation in $\mathcal{N}=4$ theory of the form considered in the previous section. Since $S(a)$ is an even function of $a$, all $\omega_{2i+1}=\frac{1}{N}\left< \tr a^{2i+1}\right>_{\ddef}=0$ and only the terms of the form $\tr a^m \tr a^n$ with even $m, \, n$ in $S(a)$ actually deform the averages.
Therefore the effective potential has the form
\begin{equation} \label{VeffN2}
V_{\eff}=2\sum_{n=2}^{\infty}(-1)^n \frac{\zeta(2n-1)}{n \, (2 \pi)^{2n}}\sum_{k=1}^{i-1} C_{2n}^{2k} \, \tr a^{2k} \omega_{2(n-k)}=\sum_{k=1}^{\infty} g_{2k} \, \tr a^{2k}
\end{equation}
with
\begin{equation} \label{g-sigma}
g_{2k}=2\sum_{n=k+1}^{\infty}(-1)^n\frac{\zeta(2n-1)}{n \, (2 \pi )^{2n}} C_{2n}^{2k} \, \tr a^{2k} \omega_{2(n-k)} \, .
\end{equation}
The system of equations for $\omega_{2n}$ following from the resolvent can be written as
\begin{equation} \label{sys-omega}
\sum_{n=1}^{\infty}\left(\delta_{k,n}-A_{k,n}\right)\frac{2^{2n}\omega_{2n}}{b^{2n}(2n)!}=\frac{b^2}{\lambda}\frac{1}{k!(k+1)!}
\end{equation}
with the matrix $A_{k,i}$ defined as
\begin{equation}
A_{k,n}=2\sum_{l=1}^{\infty}\frac{(-1)^{l+n}b^{2(l+n)}(2(1l+n))!}{(k+l)(l-1)!^2 k!^2 (n+k)}\frac{\zeta(2(n+l)-1)}{(4 \pi)^{2(n+l)}} \, .
\end{equation}
It seems that the infinite system (\ref{sys-omega}) cannot be solved in the general case, but since $b, a \sim \sqrt{\lambda}$, the system can be treated in perturbation theory in terms of a small $\lambda$.
\par For example up to $\lambda^5$ in the WL one will find the effective potential as
\begin{equation}
V_{\eff}=\frac{\tr a^2}{ \lambda}\left(\frac{3}{2^5} \, \frac{\zeta(3)}{\pi^4} \, b^4 - \frac{5}{2^8} \, \frac{\zeta(5)}{\pi^6} \, b^6 + \frac{9}{2^{11} } \, \frac{\zeta(3)^2}{\pi^8} \, b^8 +\frac{35}{2^{13}}\, \frac{\zeta(7)}{\pi^8} \, b^8\right)+
\end{equation}
$$
+ \frac{\tr a^4}{\lambda}\left( -\frac{5}{2^7} \, \frac{\zeta(5)}{\pi^6} \, b^4 +\frac{35}{2^{11}} \, \frac{\zeta(7)}{\pi^8}\, b^6  \right) + \frac{\tr a^6}{\lambda}  \frac{7}{2^9} \, \frac{\zeta(7)}{\pi^8} \, b^4  +\mathcal{O} \left(\lambda^5 \right) \, ,
$$
where $b$ should be found perturbatively from the equation
\begin{equation}
\frac{b^4}{\lambda}\left[1+\frac{3}{2^6}\, \frac{\zeta(3)}{\pi^4} \, b^4  - \frac{5}{2^9} \, \frac{\zeta(5)}{\pi^6} b^{6} -\frac{15}{2^9} \frac{\zeta(5)}{\pi^6} b^8+\frac{9}{2^{12}}\frac{\zeta(3)^2}{\pi^8}b^8 +\frac{35}{2^{14}} \frac{\zeta(7)}{\pi^8} b^8\right] +\mathcal{O}(\lambda^6)=1 \, .
\end{equation}
The WL one will get in the following form
\begin{equation} \label{WLN2expanded}
W_{\mathcal{N}=2}=W^{(0)}-\frac{3}{2}\frac{\zeta(3)}{(4 \pi)^4} \lambda^3+ \left( -\frac{1}{8}\frac{\zeta(3)}{(4 \pi)^4} +15\frac{\zeta(5)}{(4 \pi)^6} \right) \lambda^4+
\end{equation}
$$
+\left( -\frac{1}{256} \frac{\zeta(3)}{(4 \pi)^4}+\frac{65}{48} \frac{\zeta(5)}{(4 \pi)^6}+36 \, \frac{\zeta(3)^2}{(4 \pi)^8}  -140 \frac{\zeta(7)}{(4 \pi)^8}\right) \lambda^5+\mathcal{O}(\lambda^6) \, ,
$$
where $W^{(0)}$ stands for the non-perturbed WL in $\mathcal{N}=4$ theory (\ref{WLnonpert}).
\par This expansion has a simple diagrammatic interpretation. For example, the terms proportional to $ (\zeta(3))^1$ come from a diagrams with one double-vertex of the form $\left(-3 \frac{\zeta(3)}{(2 \pi)^4} (\tr a^2)^2\right)$, i.e. from the diagrams with two single two-pointed vertices, where one vertex is connected with the WL and one is disconnected (self-contracted). A self-contracted two-pointed vertex brings a factor $\frac{\lambda}{4}$ and the two-pointed vertex connected with the WL brings a factor of $ \frac{\lambda}{2} I_2(\sqrt{\lambda})$ (see (\ref{bessel})). There is also a multiplicative factor of 2 due to the two possible ways to pick up a vertex to connect with the WL. Therefore all the terms proportional to $ (\zeta(3))^1$ come from the expansion of $\left(-12 \frac{\zeta(3)}{(4 \pi)^4} \lambda^2 I_2(\sqrt{\lambda})\right)$ in any order of $\lambda$ including the ones in (\ref{WLN2expanded}).

\section{Correlators between the WL and chiral operators}

It was shown in \cite{Fucito} that the correlators between the WL and the powers of chiral operators in the fundamental representation of the non-deformed $\mathcal{N}=4$ SYM theory can also be written as $U(N)$ matrix model averages
\begin{equation}
\left< \tr \varphi^n \, \tr_{\bf R} \, e^{\mathcal{C}}\right>_{\vev}=\left<\tr (i \, a)^n \, \tr_{\bf R} \, e^{a}\right> \, ,
\end{equation}
where $\left< \,\, \right>$ stands for the averages defined in (\ref{average}) with quadratic potential $V(a)=\frac{2}{\lambda} \tr a^2$.
\par In the following subsection we are going to compare such correlators calculated on $S^4$ against the classical result \cite{Semenoff} observed on $\mathbb{R}^4$. In \cite{Gerch} the authors showed that both in $\mathcal{N}=2$ and $\mathcal{N}=4$ theories CPOs on a sphere have non-zero averages. In order to compare the average value of the CPO with the results of the gauge theory on $\mathbb{R}^4$ one should  construct new operator by subtracting from the CPO all CPOs with lower dimensions with such coefficients that all products of the new operator with anti-CPO of lower dimensions would vanish.
\par In \cite{billo} the authors compared the correlators between CPOs and anti-CPOs found in the gauge $\mathcal{N}=2$ theory on $\mathbb{R}^4$ with the correlators observed from the matrix model on $S^4$. To get rid of non-zero matrix model averages the authors used the Wick product, \textit{i.e.} they subtracted from the matrix trace $\tr a^n$ all matrix traces and products of matrix traces of lower total dimensions in such a way, that an average of the construction $:\tr a^n:$ turned to zero as well as $\left< :\tr a^n: \tr a^{k_1} \ldots \tr a^{k_m} \right>$ whenever $\sum_{i=1}^m k_i < n$.
\par In the following subsections we will also use that the Wick product on the matrix model side corresponds to the unmixing procedure for CPOs in the $\mathcal{N}=4$ theory and denote it as
\begin{equation}
\left< :\tr \varphi^n: \, \tr_{\bf R}  \, e^{\mathcal{C}}\right>_{\vev}= \left<:\tr (i \, a)^n : \, \tr_{\bf R} \,  e^{a}\right> \, ,
\end{equation}
where  $:\tr a^n :$ stands for the n-th Wick product and $:\tr \varphi^n:$ stands for the unmixed CPO.
\par We will use a definition of the Wick product equivalent to the one used in \cite{billo}.
\begin{equation} \label{norm}
:\tr a^n:  =\tr a^n-\tr a^n_1+\tr a^n_2-...+(-1)^{\left[ \frac{n}{2} \right]} \tr a^n_{\left[ \frac{n}{2} \right]} \, ,
\end{equation}
where $\tr a^n_l$ denotes $\tr a^n$ with $l$ self-contractions.

\subsection{Fundamental representation}
\par As it is clear from (\ref{Wfund}), the expectation value of the deformed WL with the potential (\ref{pot-fuc}) appears to be a generating function of connected correlators of non-normal ordered chiral primary operators with the WL in the undeformed gauge theory, in particular,
\begin{equation}
W = W^{(0)}-g \, W^{(1)}(n)+... \, ,
\end{equation}
\begin{equation}
W^{(1)}(n)=\left< \tr a^n \tr e^a \right>_c=\left< \tr a^n \, \tr e^a \right>-\left< \tr a^n\right> \left<\tr e^a \right> \, .
\end{equation}
$W^{(1)}$ counts all diagrams with one $n$-point vertex generated by the WL. As it follows from (\ref{norm}), in order to calculate the average $\left< :\tr a^n: \tr e^a \right>$, one needs to subtract from $W^{(1)}$ all contributions given by the diagrams where some of the legs coming out of the $n$-point vertex are contracted among themselves and leave only the diagrams with propagators starting from the $n$-point vertex and ending on the WL.
\par Taking into account that every propagator gives a factor $\lambda/4$, one can write
\begin{equation}
\left< \tr e^a :\tr a^n:\right>_c =\tilde{W}^{(1)}(n)=W^{(1)}(n)-\sum_{l=1}^{\frac{n}{2}-1} B^n_l \left( \frac{\lambda}{4} \right)^l \tilde{W}^{(1)}(n-2l)
\end{equation}
The coefficients $B^n_l$ are the ratios of the number of planar diagrams with $l$ self-contractions in the $n$-point vertex and $(n-2l)$ legs connecting the vertex with the WL to the number of planar diagrams with a $(n-2l)$-point vertex without self-contractions.
\par In order to find the coefficients $B^n_l$ let us introduce the quantities $K_r(m)$ standing for the number of ways in which the legs of a  $(2m+r)$-point vertex can have $m$ self-contractions and leave $r$ external legs (staying planar). With the help of Appendix A one finds
\begin{equation} \label{appendixb}
K_r(m)=\frac{(r+2m)!(r+1)}{(r+m+1)!m!} \, .
\end{equation}
So for the coefficients $B^n_l$ one finds then
\begin{equation}
B^n_l=\frac{n K_{n-2l-1}(l)}{(n-2l)}=\frac{n!}{(n-l)!l!}=C_n^l \, ,
\end{equation}
where $C_n^l$ are the binomial coefficients. Therefore
\begin{equation} \label{rec1}
W^{(1)}(n)=\sum_{l=0}^{\frac{n}{2}-1}C_n^l\left( \frac{\lambda}{4} \right)^l  \tilde{W}^{(1)}(n-2l) \, .
\end{equation}
(\ref{rec1}) can be inverted (see Appendix A), which leads to
\begin{equation} \label{rec2}
\tilde{W}^{(1)}(n)=\sum_{l=0}^{\frac{n}{2}-1}(-1)^l A^n_l \left(\frac{\lambda}{4}\right)^l W^{(1)}(n-2l) \, ,
\end{equation}
where
\begin{equation} \label{appendixc}
A^n_l=\frac{n(n-l-1)!}{l!(n-2l)!} \, .
\end{equation}
Using (\ref{rec2}) and the first term of (\ref{WLSeries})
\begin{equation} \label{W1fund}
W^{(1)}(n)=\sum_{m=1}^{\infty}\frac{C(n) \,(\frac{\lambda}{4})^{\frac{n}{2}+m}}{m! \,(m-1)! \,(\frac{n}{2}+m)}
\end{equation}
one can check that
\begin{equation} \label{bessel}
\tilde{W}^{(1)}(n)=\frac{n}{2^n}\lambda^{\frac{n}{2}}I_n(\sqrt{\lambda})
\end{equation}
which is in the agreement with the result \cite{Semenoff} and with the conjecture that the radiative corrections cancel to all orders and the sum of planar rainbow diagrams is enough to calculate the correlators. This result was also obtained in \cite{Diego}, but the general form for the coefficients $A^n_l$ was not proven.

\subsection{$K$-symmetric representation, $K/N \rightarrow 0$ limit}

In the $K$-symmetric representation the expectation value of a WL written as the matrix model integral, is
\begin{equation} \label{WsymK}
W_K=\frac{1}{N}\left<\trK e^{\mathcal{C}}\right>_{\vev}=\frac{1}{N}\left<\trK e^{a}\right>_{\ddef} \, ,
\end{equation}
where $\trK e^{a}$ stands for the WL in the $K$-symmetric representation.
\par In the large t'Hooft coupling constant limit $\lambda \gg 1$ the WL (\ref{WsymK}) can be written up to exponentially suppressed terms as \cite{Halmagyi:2007rw, Okuda:2008px}
\begin{equation}
W_K=\frac{1}{Z}\int_{-\infty}^{\infty}d a \, \Delta(a) \, e^{-NV(a)+K a_1} \, .
\end{equation}
One can write then
\begin{equation}
W_K=\frac{1}{N}\frac{1}{Z} \int_{-\infty}^{\infty}d a \, \Delta(a) \, e^{-NV(a)}\sum_{i=1}^{N}e^{K a_i}=\frac{1}{N}\frac{1}{Z}\int d[a] \tr e^{K a}e^{-NV(a)} \, ,
\end{equation}
where $\tr e^{K a}$ stands for the trace in the fundamental representation.
\par It implies that the WL in the $K$-symmetric representation has the same diagrammatic interpretation as the WL in the fundamental representation up to a factor $K$ for every line coming to the WL.
\par In particular, in the case of the potential (\ref{pot-fuc}) in the planar limit {$N \rightarrow \infty,K/N  \rightarrow 0$
\begin{equation}
W_K=\frac{1}{N}\sum_{m=0}^{\infty}\frac{K^{2m}\left< \tr a^{2m}\right>}{4^m (2m)!}=1+\sum_{m=1}^{\infty}\sum_{k=0}^{\infty}\frac{(-C(n)g_n (\frac{\lambda}{4})^\frac{n}{2})^k(\frac{\lambda K^2}{4})^m(\frac{n}{2}k+m-	1)!}{m!(m-1)!k!((\frac{n}{2}-1)k+m+1)!}
\end{equation}
can be reduced to the WL in the fundamental representation by introducing
\begin{equation}
\lambda^{new}=\lambda K^2, \quad g_n^{new}=\frac{g_n}{K^n} \, .
\end{equation}
So
\begin{equation} \label{WSWF}
W_K(\lambda, g_n)=W \left(\lambda K^2,\frac{g_n}{K^n} \right) \, .
\end{equation}
At the zeroth order it gives
\begin{equation}
W_K^{(0)}(\lambda)=W^{(0)}(\lambda K^2)=\frac{2 I_1(\sqrt{\lambda} K)}{\sqrt{\lambda} K}
\end{equation}
which recovers the known result \cite{Okuyama:2006jc} and in the first order
\begin{equation} \label{ratio}
W_{K}^{(1)}(n,\lambda)=\frac{1}{K^n} W^{(1)}(n,\lambda K^2) \, .
\end{equation}
Using (\ref{ratio}) and (\ref{rec1}) one can write that
$$
W_{K}^{(1)}(n,\lambda)=\frac{1}{K^n} \sum_{l=0}^{\frac{n}{2}-1}C_n^l \left( \frac{\lambda}{4} \right)^l K^{2l}\tilde{W}^{(1)}(n-2l,\lambda K^2)=
$$
\begin{equation} \label{corrsmallK}
= \sum_{l=0}^{\frac{n}{2}-1}C_n^{2l} \left( \frac{\lambda}{4} \right)^l  \frac{\tilde{W}^{(1)}(n-2l,\lambda K^2)}{K^{n-2l}}=\sum_{l=0}^{\frac{n}{2}-1}C_n^l \left( \frac{\lambda}{4} \right)^l  \tilde{W}_{K}^{(1)}(n-2l,\lambda) \, ,
\end{equation}
so the recurrent expression (\ref{rec1}) holds.
\par An explicit expression for $\tilde{W}_{K}^{(1)}(n,\lambda)$ is
\begin{equation} \label{W1K/N0}
\tilde{W}_{K}^{(1)}(\lambda)=\frac{n}{2^n}\lambda^{\frac{n}{2}}I_n(K\sqrt{\lambda}) \, .
\end{equation}

\subsection{$K$-symmetric representation, finite $K/N$ limit}
	
In the case of the $K$-symmetric representation with $K$ behaving as $N$ the expectation value of the WL or its correlators with primary operators can be no more directly expressed in terms of planar diagrams since these quantities grow exponentially with $N$ and so their expansion in large-$N$ series makes no sense \cite{MFS}.
\par (\ref{rec1}), (\ref{rec2}) were derived by a direct counting of planar diagrams but nevertheless it can be shown that they still hold! In order to do that a planar diagrammatic interpretation can be restored (see Appendix B).
\par To prove (\ref{rec2}) in the limit of $K \sim N$ let us look at the definition of the n-th Wick product (\ref{norm}).
\par The key point of the proof is that as it is shown in the Appendix B both for $K \ll N$ and $K \sim N$ one can write in the highest order of the large-$N$ expansion
\begin{equation} \label{planar}
\left<  \tr a^{n_1}...\tr a^{n_l} \trK e^{a}\right>_c=\lambda^{\frac{n_1+...+n_l}{2}}N^{l-1}\sum_{i=1}^{l} \alpha_i \frac{\left< \tr a^{n_i}\trK e^{a}\right>_c}{\lambda^{\frac{n_i}{2}}} \, ,
\end{equation}
where $\alpha_i$ are numerical coefficients not depending on $N$, $K$ or $\lambda$.
	\par Furthermore, using the relations for the generators of an $U(N)$ group one sees that the quantity $\left<  \tr a^n_l \trK e^{a}\right>_c$ can be expressed as a sum of correlators of the WL with products of traces of lower order $\left<  \tr a^{n_1}...\tr a^{n_m} \trK e^{a}\right>_c$, $n_1+...n_{m}=n-2l$, $m \in [1,l+1]$. According to (\ref{planar}) at the highest order in $N$ only terms with maximal number of traces $m=l+1$ contribute in $\left<  \tr a^n_l \trK e^{a}\right>_c$, so one can write
\begin{equation} \label{l1}
\left<  \tr a^n_l \trK e^{a}\right>_c=\frac{\lambda^l}{N^l} \sum_{k_1+...+k_{l+1}=n-2l} \beta(k_1,...,k_{l+1})\left< \tr a^{k_1}...\tr a^{k_{l+1}} \trK e^{a}\right>_c \, ,
\end{equation}
where $\beta(k_1,...,k_{l+1})$ are combinatoric coefficients not depending on $\lambda$, $K$ or $N$.
\par From (\ref{planar}) and (\ref{l1}) one gets
$$
\left<  \tr a^n_l \trK e^{a}\right>_c=\frac{\lambda^l}{N^l}\lambda^{\frac{n}{2}-l} \sum_{k_1+...+k_{l+1}=n-2l} \beta(k_1,...,k_{l+1})N^{l}\sum_{i=1}^{l+1} \alpha_i \frac{\left< \tr a^{k_i}\trK e^{a}\right>_c}{\lambda^{\frac{k_i}{2}}}=
$$
\begin{equation} \label{lconv}
=\lambda^{\frac{n}{2}} \sum \gamma_{k_i} \frac{\left< \tr a^{k_i}\trK e^{a}\right>_c}{\lambda^{\frac{k_i}{2}}} \, ,
\end{equation}
where $\gamma_{k_i}$ are coefficients not depending on $\lambda$, $K$ or $N$.
From (\ref{norm}) and (\ref{lconv}) the expression (\ref{rec2}) follows
$$
\left<  :\tr a^n: \, \trK e^{a}\right>_c=\sum_{l=0}^{\frac{n}{2}-1} (-1)^l \left( \frac{\lambda}{4} \right)^l A^n_l \left< \trK e^{a} \, \tr a^{n-2l} \right>_c \, .
$$
The coefficients $A^n_l$ were found earlier for the case $K \ll N$. From the reasoning above it is clear that (\ref{rec2}) holds in the case $K \sim N$ as well with the same coefficients $A^n_l$.
\par Beside this in the Appendix B with the help of restored planar diagrammatic approach it is shown that
\begin{equation} \label{W1/NW}
\frac{1}{N}\frac{W_{K}^{(1)}}{W_{K}^{(0)}}=\frac{\left<  \tr a^n \, \trK e^{a}\right>_c}{\left<  \trK e^{a}\right>}=\lambda^{\frac{n}{2}} (1+y^2)^{\frac{n}{2}} -\frac{\lambda^{\frac{n}{2}}}{2^n}\frac{n!}{(\frac{n}{2})!(\frac{n}{2})!}
\end{equation}
$$
-\frac{\lambda^{\frac{n}{2}}}{2^n}\frac{(n)!}{(\frac{n}{2})!(\frac{n}{2}-1)!}\sum_{m=1}^{\infty}\frac{1}{m 2^{2m}(1+y^2)^m}\frac{(2m)!}{m! (m-1)!}\frac{1}{\frac{n}{2}+m} \, ,
$$
where $y=K\sqrt{\lambda}/4N$.
\par Applying now (\ref{rec2}) to (\ref{W1/NW}) one finds that for an arbitrary $n$ 
\begin{equation} \label{W1norm/NW}
\frac{1}{N}\frac{\tilde{W}_{K}^{(1)}}{W_{K}^{(0)}}=2^{1-n}\sinh(n \sinh^{-1}(y))
\end{equation}
which is in the agreement with \cite{Giombi} up to a normalization.
\par It should be remarked that the results from previous subsection cannot be found as the $K/N \rightarrow 0$ limit of (\ref{W1/NW}), (\ref{W1norm/NW}), since in the derivation we have crucially used the fact that $K/N$ is finite.

\section{Short summary}

In this paper we explored the connection between the vev of Wilson loops and its correlators with primary fields in the deformed SYM field theory leading to a deformed $U(N)$ matrix model. We treated the latter in the limit of large number of colours $N$.
\par For the potential of the form
\begin{equation*}
V=\sum_{l=1}^{\infty}g_l \tr a^l
\end{equation*}
we found the WL as
\begin{equation*}
W=\frac{e^c}{2 \pi i}\oint dz \, \omega(z) e^{z-c}dz=\frac{1}{2} e^{c}\sum_{l=2}^{\infty} \frac{g_l c^{l-2} b\,l!}{(l-2)!} I_1(b)+
\end{equation*}
$$
+\frac{1}{4} e^{c} \sum_{l=3}^{\infty}\sum_{k=0}^{\left[ \frac{l+1}{2} \right]-2} \frac{ g_l c^{l-3}b^2  \, l!}{(l-3-2k)!}\frac{(b/c)^{2k}}{k!(k+1)!} \left( 1 +\frac{1}{2}\frac{b}{c}\frac{(l-3-2k)}{(k+1)}\frac{\dd}{\dd b}\right)\left(\frac{1}{b}\frac{\dd}{\dd b} \right)^{k} b^{k} I_{k+2}(b) \, ,
$$
where the parameters $b$ and $c$ are real roots of the equations
\begin{equation*}
\sum_{l=2}^{\infty} g_l c^{l-2} b^2 \sum_{k=0}^{\left[ \frac{l}{2}\right]-1}\frac{1}{2^{2k+2}}\frac{l!}{(l-2-2k)!}\frac{(b/c)^{2k}}{k!(k+1)!}-1=0 \, ,
\end{equation*}
\begin{equation*}
\sum_{l=1}^{\infty}g_l c^{l-1} \sum_{k=0}^{\left[\frac{l+1}{2}\right] -1} \frac{1}{2^{2k}}\frac{l!}{(l-1-2k)!}\frac{(b/c)^{2k}}{k!^2}=0 \, .
\end{equation*}
We showed that the case of the potentials of the form
\begin{equation*}
V(a)=\frac{2}{\lambda} \tr a^2 + \frac{g}{N} \tr a^n \tr a^m
\end{equation*}
can be reduced to the previous one with the effective potential
\begin{equation*}
V_{\eff}=\frac{2}{\lambda} \tr a^2 + g \, \omega_m \tr a^n + g \, \omega_n \tr a^m 
\, ,
\end{equation*}
where the parameters $\omega_m, \, \omega_n$ are defined by the resolvent (\ref{res-arb-def}).
\par We considered  also the WL in the $\mathcal{N}=2$ SYM theory as the deformed WL in the $\mathcal{N}=4$ theory and found that
\begin{equation*}
W_{\mathcal{N}=2}=W^{(0)}-\frac{3}{2}\frac{\zeta(3)}{(4 \pi)^4} \lambda^3+ \left( -\frac{1}{8}\frac{\zeta(3)}{(4 \pi)^4} +15\frac{\zeta(5)}{(4 \pi)^6} \right) \lambda^4+
\end{equation*}
$$
+\left( -\frac{1}{256} \frac{\zeta(3)}{(4 \pi)^4}+\frac{65}{48} \frac{\zeta(5)}{(4 \pi)^6}+36 \, \frac{\zeta(3)^2}{(4 \pi)^8}  -140 \frac{\zeta(7)}{(4 \pi)^8}\right) \lambda^5+\mathcal{O}(\lambda^6) \, ,
$$
where $W^{(0)}$ stands for the non-perturbed WL in $\mathcal{N}=4$ theory (\ref{WLnonpert}).
\par In the last section we considered the correlators of a WL with CPOs and found the connection between the correlators $W^{(1)}(n)$ obtained from the matrix model and the correlators with unmixed operators $\tilde{W}^{(1)}(n)$.
\begin{equation*}
W^{(1)}(n)=\sum_{l=0}^{\frac{n}{2}-1}C_n^l \left( \frac{\lambda}{4} \right)^l \tilde{W}^{(1)}(n-2l) \, ,
\end{equation*}
\begin{equation*}
\tilde{W}^{(1)}(n)=\sum_{l=0}^{\frac{n}{2}-1}(-1)^l A^n_l \left( \frac{\lambda}{4} \right)^l  W^{(1)}(n-2l), \,\, A^n_l=\frac{n(n-l-1)!}{l!(n-2l)!} \, .
\end{equation*}
We showed that these relations hold not only in the fundamental representation, but also in a symmetric one. These relations were found by counting planar diagrams, although the latter are not enough to describe the WL and its correlators in the $K$-symmetric representation with $K$ behaving as $N$.

\acknowledgments
I would like to thank F.Fucito and J.F.Morales for having suggested the problem and for constant encouragment and advice during the completion of this work.
\par This work  is partially supported by the MIUR PRIN Contract 2015MP2CX4 ``Non-perturbative Aspects Of Gauge Theories And Strings''.

\appendix

\section{Some combinatoric calculations}

\subsection{Coefficients $P(n)$}
Let us introduce a quantity $P(n)$ giving the number of ways to couple in a planar way $2n$ lines which start at the same $2n$-point vertex staying in one half-plane.
\par Summing over all possible ways to couple the very left leg one can notice that
\begin{equation}
P(n)=\sum_{k=0}^{\frac{n}{2}-1}P(k)P(n-1-k) \, .
\end{equation}
It implies an equation for the generating function $f(p)=\sum_{m=0}^{\infty}p^m P(m)$
\begin{equation}
pf(p)^2-f(p)+1=0 \, ,
\end{equation}
so
\begin{equation}
f(p)=\frac{1-\sqrt{1-4p}}{2p}
\end{equation}
and therefore
\begin{equation}
P(m)=\frac{(2m)!}{m!(m+1)!} \, .
\end{equation}

\subsection{Coefficients $K_r(m)$}
To find $K_r(m)$ which is the number of ways to do $m$ couplings and leave $r$ external legs in a planar way in a $(2m+r)$-point vertex staying in the half-plan one can sum over all possible positions of the very left coupled leg and then over the number of pairs of leg isolated by the loop formed with this first leg and write that
\begin{equation}
K_r(m)=\sum_{t=0}^{r}\sum_{l=0}^{m-1}P(l)K_{r-t}(m-1-l) \, .
\end{equation}
Hence the generating function $g(p,q)=\sum_{m,r=0}^{\infty}p^mq^rK_r(m)$ has the following form
\begin{equation}
g(p,q)=\frac{1}{1-q-pf(p)}=\frac{2}{1-2q+\sqrt{1-4p}} \, .
\end{equation}
Expanding it in a series over $q$
\begin{equation}
g(p,q)=f(p)+q f(p)^2+q^2 f(p)^3+...
\end{equation}
one sees that
\begin{equation}
f(p)^{r+1}=\sum_{m=0}^{\infty} K_r(m)p^m \, .
\end{equation}
So the needed quantities can be expressed as
\begin{equation}
K_r(m)=\frac{1}{2\pi i}\oint d \zeta \zeta^{2m-1}f\left( \frac{1}{\zeta^2}\right)^{r+1}=\frac{(r+2m)!(r+1)}{(r+m+1)!m!}
\end{equation}

\subsection{Coefficients $A_l^n$}
To find the coefficients $A^n_l$ in (\ref{rec2}) one can substitute $W^{(1)}(n)$ from (\ref{rec1}) into (\ref{rec2}) and observe that
\begin{equation}
\sum_{k=0}^{s}(-1)^k A^n_{s-k}C_{n-2s+2k}^k=\delta_{s,0} \, .
\end{equation}
One can write then
\begin{equation}
\sum_{s=0}^{\frac{n}{2}-1}(x^s+x^{n-s})\sum_{k=0}^{s}(-1)^k A^n_{s-k}C_{n-2s+2k}^k=(1+x^n) \, .
\end{equation}
Using that one derives an equation for the generating polynomial $h(x)=\sum_{k=0}^{\frac{n}{2}-1}A_k^ny^k$
\begin{equation}
h \left(\frac{x}{(1-x)^2}\right)(1-x)^n - x^{\frac{n}{2}}\zeta=1+x^n \, ,
\end{equation}
where $\zeta=\sum_{\delta=0}^{\frac{n}{2}-1}A^n_{\delta}(-1)^{\delta+\frac{n}{2}}C_{n-2 \delta}^{\frac{n}{2}-\delta}$ is a quantity not depending on $x$.
\par Taking into account that $A^n_0=1$ one gets for the generating polynomial
\begin{equation}
h(y)=y^{\frac{n}{2}}\zeta + y^n j^n(y) + j^{-n}(y) \, ,
\end{equation}
where $y=\frac{x}{(1-x)^2}$ and $j(y)=\frac{-1+\sqrt{1+4y}}{2y}$. The first two terms cancel against the higher orders in the polynomial $h(y)$, so all the information about $A^n_l$ is in the third term. Finally one can find the needed coefficients by contour integration
\begin{equation}
A_k^n=\frac{1}{2 \pi i }\oint z^{2k-1} (-1)^k j^{-n}\left(-\frac{1}{z^2}\right) dz=\frac{n(n-k-1)!}{k!(n-2k)!} \, .
\end{equation}

\subsection{Coefficients $S(n, \,m)$}
Another combinatoric coefficient which we use in Appendix B is the number of two-vertices planar diagrams with one n-point vertex and one k-point vertex $S(n,k)$.
\par To find $S(n,k)$ one can notice that the first leg starting in the n-point vertex either couples with a leg starting in the k-point vertex or returns to the same vertex. Hence one can write a relation
\begin{equation}
S(n, \, m)=m \, P\left(\frac{n+m-2}{2} \right)+2 \sum_{\delta=0}^{\frac{n-3}{2}}P(\delta)S(n-2(1+\delta),\,m)
\end{equation}
and for the generating function $k_m(x)=\sum_{t=1}^{\infty}x^t S(t, \, m)$ one gets
\begin{equation}
k_m(x)=\frac{m x^2}{x^m\left(1-2x^2f(x^2)\right)}\left(f(x^2)-\sum_{t=0}^{\frac{k-3}{2}} x^{2t} P(t) \right) \, .
\end{equation}
Using that $S(n,\, m)=\frac{1}{2 \pi i }\oint \frac{dz}{z^{n+1}}k_m(z)$ and some algebra one can find $S(n,\, m)$ in a symmetric form
\begin{equation}
S(n,\, m)=\frac{n! \, m!}{\left[\frac{n}{2}\right]!\left[\frac{n-1}{2}\right]!\left[\frac{m}{2}\right]!\left[\frac{m-1}{2}\right]!} \frac{1}{\left( \frac{n+m}{2}\right)} \, .
\end{equation}

\section{Proof of (\ref{planar})}

\par In order to get back the planar diagrammatic interpretation of the WL and the correlators and prove (\ref{planar}) in the case $K \sim N$ let's consider the WL in the $K$-symmetric representation as an integral over hermitian matrices
\begin{equation}
W_K^{(0)}= \frac{1}{N} \frac{\int d[a] \trK e^a e^{-\frac{2 N}{\lambda} \tr a^2}}{\int d[a] e^{-\frac{2N}{\lambda} \tr a^2}}=\frac{1}{N}\int d\mu[v] \trK e^{v\frac{\sqrt{\lambda}}{2\sqrt{N}}} \, ,
\end{equation}
where $d[a]$ is the Lebesgue measure in the space of hermitian $N\times N$ matrices, $v=\frac{\sqrt{N}}{\sqrt{\lambda}}a$ and $d\mu[v]$ is the Gaussian measure in the space of hermitian matrices of size $N\times N$
$$d\mu[v]= 2^{\frac{N^2-N}{2}} \left( 2\pi\right)^{-\frac{N^2}{2}}e^{-\frac{1}{2}tr v^2} d[v], \,\, \int d\mu[v]=1 \, .$$
In the large $\lambda$ limit one can rewrite WL as \cite{Okuda:2008px}, \cite{lando}
\begin{equation} \label{WLexact}
W_K^{(0)}=\frac{1}{N}c_N \int \prod_{i=0}^{N-1} d\mu[v_i] \prod_{u<w=0}^{N-1}(v_u-v_w)^2 N e^{\frac{K v_0}{2}\frac{\sqrt{\lambda}}{\sqrt{N}}} \, ,
\end{equation}
where $d\mu[v_i]$ is the Gaussian measure in the space of real eigenvalues
$$
d\mu[v_i]=\frac{1}{\sqrt{2\pi}}e^{-\frac{v_i^2}{2}}dv_i
$$
and
$$
c_N=\frac{\vol \left( U(N) \right)}{N!\left( 2\pi\right)^{\frac{N^2+N}{2}}}=\frac{1}{N! \prod_{k=1}^{N-1}k!} \, .
$$
\par Now one can separate an integral over $v_0$ with the large exponent $e^{\frac{K v_0}{2}\frac{\sqrt{\lambda}}{\sqrt{N}}}$.
$$
W_K^{(0)}=c_N \int d\mu[v_0]e^{\frac{K v_0}{2}\frac{\sqrt{\lambda}}{\sqrt{N}}}\int \prod_{i=1}^{N-1} d\mu[v_i] \prod_{u<w=1}^{N-1}(v_u-v_w)^2\prod_{k=1}^{N-1}(v_0-v_k)^2=
$$
$$
=c_N \int d\mu[v_0]e^{\frac{K v_0}{2}\frac{\sqrt{\lambda}}{\sqrt{N}}}\int \prod_{i=1}^{N-1} d\mu[v_i] \prod_{u<w=1}^{N-1}(v_u-v_w)^2e^{2\sum_{k=1}^{N-1}\ln\left(v_0-v_k\right)}=
$$
$$
=\frac{c_N}{c_{N-1}} \int d\mu[v_0]e^{\frac{K v_0}{2}\frac{\sqrt{\lambda}}{\sqrt{N}}}\int d\mu[\tilde{v}]e^{2\, \tr\left( \ln(v_0 - \tilde{v})\right)} \, ,
$$
where $\tilde{v}$ is a hermitian $(N-1) \times (N-1)$ matrix.
$$
W_K^{(0)}=\frac{1}{N!} \int d\mu[v_0]e^{\frac{K v_0}{2}\frac{\sqrt{\lambda}}{\sqrt{N}}}\int d\mu[\tilde{v}]e^{2\, \tr\left( \ln(v_0 - \tilde{v})\right)}=
$$
\begin{equation} \label{WLdoubleint}
=\frac{1}{N!} \int d\mu[v_0]e^{\frac{K v_0}{2}\frac{\sqrt{\lambda}}{\sqrt{N}}}v_0^{2(N-1)}\int d\mu[\tilde{v}]e^{2\, \tr\left( \ln(1- \frac{\tilde{v}}{v_0})\right)} \, .
\end{equation}
Let us now take an internal integral and consider the exponent as a new interaction. Let us also assume for now that $v_0$ is greater than any eigenvalue of $\tilde{v}$.
\begin{equation}
C=\int d\mu[\tilde{v}]e^{-2 \, \tr\sum_{k=1}^{N-1}\frac{\tilde{v}^k}{v_0^k}\frac{1}{k}} 
\end{equation}
Perturbation theory shows that every propagator gives a factor 1 and the $k$-point Vandermonde vertex gives a factor
$$-2\frac{1}{v_0^k k}$$
One needs to take into account both connected and disconnected diagrams. Both contributions can be expressed as an exponent of the sum over connected diagrams. To recover the correct exponential behavior in the large $N$ limit only one vertex diagrams can be taken in the sum. Indeed, in this case every cycle brings the factor $(N-1) \approx N$ and propagators and the vertices don't bring the factor $N$, so the total power of $N$ is proportional to the number of faces of the planar diagram $F=2+L-V$ and we need to take the minimal nontrivial number of vertices $V$ for every number of lines $L$. Therefore
\begin{equation} \label{exp}
C=\exp\left(-\sum_{k=1}^{\infty} \frac{P(k)}{k}\frac{N^{k+1}}{ v_0^{2k}} + \mathcal{O}(1) \right) \, .
\end{equation}
Taking the sum one gets
\begin{equation} \label{C}
\ln(C)=-\frac{v_0^2}{2}\left(\sqrt{1-\frac{4N}{v_0^2}}-1\right)-N\left(1+\ln4\right)+2N\ln\left(1+\sqrt{1-\frac{4N}{v_0^2}}\right) +\mathcal{O}(1) \, .
\end{equation}
Substituting now (\ref{C}) into (\ref{WLdoubleint}) one can take the integral over $v_0$ in the large $N$ limit with the saddle point method getting
\begin{equation} \label{SadP}
v_0=2\sqrt{N}\sqrt{1+y^2} \, ,
\end{equation}
where $y=\frac{K \sqrt{\lambda}}{4N}$. The saddle point is greater than any eigenvalue of the matrix $\tilde{v}$ belonging to the interval $[-2\sqrt{N},2\sqrt{N}]$ if $y > 0$. So the approach doesn't work for $K/N \rightarrow 0$, but only for finite $K/N$.
\par Hence we get
\begin{equation} \label{WLresult}
W_K^{(0)} \approx \frac{1}{\sqrt{2\pi N}}\left( \frac{y^2}{1+y^2} \right)^{\frac{1}{4}}\frac{1}{N\left(2\sqrt{1+y^2}\right)^2}e^{-N}e^{2N\left(y\sqrt{1+y^2}+\sinh^{-1}(y)\right) + \mathcal{O}(1)} \, .
\end{equation}
The non-deformed WL in $K$-symmetric representation with finite $N$ can be written as \cite{MFS}
\begin{equation}
W_K^{(0)}=\frac{1}{N} e^{\frac{K^2 \lambda}{8 N}}L_{N-1}^1\left( -\frac{K^2 \lambda}{4 N}\right)
\end{equation}
and the large $N$ limit of it is
\begin{equation}
W_K^{(0)} \approx \frac{1}{\sqrt{2\pi N}}\left( \frac{y^2}{1+y^2} \right)^{\frac{1}{4}}\frac{1}{N\left(2y\right)^2}e^{-N}e^{2N\left(y\sqrt{1+y^2}+\sinh^{-1}(y)\right)} \cdot \left(1+ \mathcal{O}\left(\frac{1}{N}\right) \right) \, .
\end{equation}
Which coincide with (\ref{WLresult}) up to a pre-exponential factor. To recover it one needs to calculate the next order in the large-$N$ expansion in the power of the exponent in (\ref{WLresult}).
\par Let us do so. There are two different contributions at the next order in $N$. The first contribution arises from the factor $N-1$ brought by every cycle. This factor cannot be simply replaced with $N$ anymore, but with $(N-1)^{n} \approx N^{n}-nN^{n-1}$ for $n$ cycles.  The second contribution is given by the two vertices planar diagrams. It should be noted that at this order in the large-$N$ expansion there is no need to take into account non-planar diagrams since 1 is the Euler characteristic of a non-orientable surface.
\par Therefore one gets the following power of the exponent (\ref{exp}) in the two highest orders in the large-$N$ expansion
\begin{equation} \label{exp2}
\ln(C)=-\sum_{k=1}^{\infty} \frac{P(k)}{k}\frac{N^{k+1}}{ v_0^{2k}} +\sum_{k=1}^{\infty} \frac{(k+1)}{k}P(k)\frac{N^{k}}{v_0^{2k}} +
\end{equation}
$$
+\frac{1}{2}\sum_{l=1}^{\infty}\sum_{m=1}^{\infty}\frac{S(2m,2l)}{l \, m}\frac{N^{l+m}}{v_0^{2(l+m)}}+\frac{1}{2}\sum_{l=0}^{\infty}\sum_{m=0}^{\infty}\frac{S(2m+1,2l+1)}{\left(l+\frac{1}{2}\right)\left(m+\frac{1}{2}\right)}\frac{N^{l+m+1}}{v_0^{2(l+m+1)}} +\mathcal{O}\left( \frac{1}{N}\right) \, ,
$$
where $S(n,k)$ is the number of two-vertices planar diagrams with one n-point vertex and one k-point vertex (see Appendix A).
\begin{equation}
S(n,k)=\frac{n!k!}{\left[\frac{n}{2} \right]!\left[\frac{n-1}{2} \right]!\left[ \frac{k}{2}\right]!\left[ \frac{k-1}{2}\right]!\left( \frac{n+k}{2}  \right)} \, ,
\end{equation}
The double sums in (\ref{exp2}) can be rewritten as a derivative of a product of two separate sums, which in their turn can be calculated. So it can be checked that in the saddle point (\ref{SadP}) three terms of the sub-leading order give
\begin{equation}
\ln(C)=-\sum_{k=1}^{\infty} \frac{P(k)}{k}\frac{N^{k+1}}{ v_0^{2k}} +\ln \left(\frac{1+y^2}{y^2} \right)+\mathcal{O}\left(\frac{1}{N}\right)
\end{equation}
which brings exactly the necessary factor to (\ref{WLresult}).
\par In order to calculate the connected correlators $\left< \tr_K e^{\frac{a}{2}} \tr a^n \right>_c$ acting in the same way as described above one can write that
$$
W_K^{(1)}=\left< \trK e^{a} \, \tr a^n \right>_c=\left< \trK e^{a}\left(a_0^n+\tr \tilde{a}^n\right)\right>-\left< \trK e^{a}\right>\left< \tr a^n\right>=
$$
$$
=\left< \trK e^{a} \, a_0^n\right>+\left< \trK e^{a} \, \tr\tilde{a}^n\right>-\left< \trK e^{a}\right>\left< \tr a^n\right> \, .
$$
It can be noticed that in the large $N$ limit only diagrams without or with one Vandermonde vertex connected with the n-point vertex should be taken into account for the same reason as above, so
$$
\left< \trK e^{a}\tr a^n \right>_c=\left< \trK e^{a} a_0^n\right>+\left< \trK e^{a} \tr \tilde{a}^n \right>'+\left< \trK e^{a}\right>\left< \tr \tilde{a}^n\right>-\left< \trK e^{a}\right>\left< \tr a^n\right> \, ,
$$
where $\left< \trK e^{a} \tr \tilde{a}^n \right>'$ stands for the diagrams where only one Vandermonde vertex is connected with the n-point vertex.
\par In the large $N$ limit
$$
\left< \tr a^n\right> = N \lambda^{\frac{n}{2}}P\left(\frac{n}{2} \right), \,\,\, \left< \tr\tilde{a}^n\right> = \frac{(N-1)^{\frac{n}{2}+1}}{N^{\frac{n}{2}}}\lambda^{\frac{n}{2}}P\left(\frac{n}{2} \right) \, ,
$$
so
$$
\left< \tr\tilde{a}^n\right> \approx \left(1-\frac{\frac{n}{2}+1}{N} \right)\left< \tr a^n\right> \, .
$$
Therefore after the change of variables $a=\frac{\sqrt{\lambda}}{\sqrt{N}}v$ one can see that an additional factor comparing with (\ref{WLdoubleint}) appears in the integral over $v_0$ which has the form
\begin{equation} \label{fact}
\frac{1}{N}\frac{W_{K}^{(1)}}{W_K^{(0)}}=\lambda^{\frac{n}{2}}\frac{v_0^n}{N^{\frac{n}{2}}} +\lambda^{\frac{n}{2}}\sum_{m=1}^{\infty}\frac{(-1) N^m}{m \, v_0^{2m}}M(n,m)-\lambda^{\frac{n}{2}}\left(\frac{n}{2}+1\right)P\left( \frac{n}{2}\right) \, ,
\end{equation}
where $M(n,m)$ is a number of planar diagrams with $2m$ lines connecting the $n$-point vertex and one Vandermonde vertex and can be taken from the resolvent (\ref{ResSeries}).
$$
M(n,m)=\frac{C(n)(2m)!}{m! (m-1)!}\frac{1}{\frac{n}{2}+m} \, .
$$
Substituting the saddle point (\ref{SadP}) into (\ref{fact}) one will find the ratio of the correlator and the WL (\ref{W1/NW}).
\par Finally, to calculate $\left< \tr a^{n_1}...\tr a^{n_l} \trK e^{a}\right>_c$ in the highest order of $N$ one can write
$$
\left< \tr a^{n_1}...\tr a^{n_l} \trK e^{a}\right>_c=\left< (a_0^{n_1}+\tr \tilde{a}^{n_1})...(a_0^{n_l}+\tr \tilde{a}^{n_l}) \trK e^{a}\right>_c=
$$
$$
=\sum_{i=1}^{l}\prod_{\stackrel{j=1}{j \neq i}}^{l}\left< \tr a^{n_j} \right>\left( \left< \trK e^{a} a_0^{n_i}\right>+\left< \trK e^{a} \tr \tilde{a}^{n_i} \right>'-\left(\frac{n}{2}+1\right)\left< \tr a^{n_i} \right>\right)=
$$
\begin{equation} \label{manytr}
=\sum_{i=1}^{l}\left< \trK e^{a} \tr a^{n_i} \right>_c\prod_{\stackrel{j=1}{j \neq i}}^{l}\left< \tr a^{n_j} \right>=\lambda^{\frac{n_1+...+n_l}{2}}N^{l-1}\sum_{i=1}^{l} \prod_{\stackrel{j=1}{j \neq i}}^{l} P\left(\frac{n_j}{2} \right) \frac{\left< \tr a^{n_i}\trK e^{a}\right>_c}{\lambda^{\frac{n_i}{2}}} \, .
\end{equation}
In the case of $K \ll N$ the planar diagrammatic interpretation is applicable to the WL and its correlators, so one can write (\ref{manytr}) immediately. The statement (\ref{planar}) is proven.

\end{document}